\documentstyle[]{elsart}
\newif\iffigure
\newif\iffigend
\figuretrue	

\chardef\atcode=\catcode`\@
\catcode`\@=11	
\iffigure
  \input rotate.sty
  \input pstricks.sty
  \input epsf.sty
  \def\@journal{Heavy Ion Physics}
  \def\ps@copyright{\def\@oddfoot{\small\sl
   GSI-Preprint 95-30, subm. to Heavy Ion Physics\hfil}}
\fi
\def\nuc#1#2{\relax\ifmmode{}^{#1}{\protect\text{#2}}\else${}^{#1}$#2\fi}
\catcode`\@=\atcode	

\newcommand {\NP}[1]{ Nucl. Phys. {\bf #1}}
\newcommand {\ZP}[1]{ Z. Phys. {\bf #1}}
\newcommand {\PR}[1]{ Phys. Rev. {\bf #1}}
\newcommand {\PRL}[1]{ {\it Phys. Rev. Lett.} {\bf #1}}

\newcommand{\PRP}[1]{ Phys. Rep. {\bf #1}}

\iffigend
  
\fi

\newcounter{fig}
\newenvironment{fig}[1]
  {\iffigure \iffigend \begin{figure}[p] {#1}
	     \else \begin{figure} {#1} \fi
             	   \refstepcounter{fig}
   \else \refstepcounter{fig} \fi}
  {\iffigure \end{figure} \fi}

\begin{document}
\begin{frontmatter}
\title{The Path of Hot Nuclei Towards Multifragmentation}
\author[GSI]{G.~Papp\thanksref{ELTE}} and
\author[GSI,TH]{W.~N\"orenberg}
\address[GSI]{Gesellschaft f\"ur Schwerionenforschung,
D-64220, Darmstadt, Germany}
\address[TH]{Institut f\"ur Kernphysik, TH Darmstadt,
D-64289, Darmstadt, Germany}
\thanks[ELTE]{On leave from the Institute for Theoretical
Physics of E\"otv\"os University, H-1088 Budapest, Hungary}

\vspace*{2truecm}

\begin{abstract}
The initial production and dynamical expansion of hot spherical nuclei are
examined as the
first stage in the projectile-multifragmentation process. The initial
temperatures, which are necessary for entering the adiabatic spinodal region,
as well as the minimum temperatures and densities, which are reached in the
expansion,
significantly differ for hard and soft equations of
state. Additional initial compression, occurring in central collisions
leads most likely to a qualitatively different multifragmentation
mechanism. Recent experimental data are discussed in relation to the results of
the proposed model.
\end{abstract}
\end{frontmatter}



\newpage

\section{Introduction}

New results \cite{kn:hub,kn:poch94} on projectile multifragmentation in
relativistic heavy-ion collisions indicate relations to the liquid-gas
phase transition of nuclear matter. Motivated by these experiments we
like to substantiate this conjecture by studying the evolution of the
excited projectile residue into the region of instability.

In the following we report results on the
expansion dynamics of hot spherical nuclei. We estimate the initial
excitation energy from the abrasion model \cite{kn:oli,kn:abr} using
modifications introduced in \cite{kn:abr1}. The dynamical expansion of the
heated spherical residue is calculated taking the evaporation of nucleons
into account~\cite{kn:he}.

The expansion of the projectile residue can be regarded as the first
stage in the multifragmentation process. For large enough initial
excitation energies and temperatures the nuclear matter of the residue
reaches states
of mechanical volume instability, where small density fluctuations grow
exponentially \cite{kn:bur92,kn:bat93,kn:pg}. For an illustrative analysis
of the
instability evolution we refer to ref. \cite{kn:col94}. The fragmentation
process itself has been studied within molecular dynamics
\cite{kn:pand85,kn:pand867,kn:paul91,kn:dor94}.
Studies by Friedman \cite{kn:fri1,kn:fri2}
within an expanding emitting source model (EES) show that intermediate
mass fragments (IMFs) are indeed created within time intervals of about
50 fm/c indicating a simultaneous breakup of the residue.

\section{The Model}

In this section we describe a few essential features of the model.
A more detailed description of the model will be published elsewhere
\cite{kn:pg2}.

We start out from a thermalized spherically symmetric projectile-like nucleus
at
ground state density, which has been created in the peripheral high-energy
collision (cf. \cite{kn:baue94}). The typical temperature for
such a hot nucleus is not too high, such that
the mean free path of nucleons is still comparable or even larger than the size
of the system \cite{kn:he}.
Thus, we can approximately assume homogeneity in density and temperature for
the expanding nucleus.

During the expansion the hot nucleus evaporates particles. We consider only
neutron and proton evaporation. Deuterons may be considered to arise from
coalescence of nucleons.

If the excitation energy is high enough, the expansion leads into the
instability region of the equation of state. For not too high excitation
energies the expansion of the homogeneous system stops at a certain density.
Around these
turning points the system has enough time ($\ge$ 30 fm/c,
\cite{kn:bur92,kn:pg}) to develop inhomogeneities in density, and subsequently
decay into many fragments.
The final mass and charge spectra is not given by this model, but
the calculated turning-point parameters (temperature and density) can be
applied to static statistical models \cite{kn:gros,kn:bond}
and related to experimental results (cf. sect. 3). In a complete dynamical
treatment of multifragmentation the turning points are the initial states for
the decay into the final fragments.

\subsection{Abrasion}

The starting point in our calculations is the determination of the
projectile residue (prefragment) which is produced in the heavy-ion collision.
A simple
picture for relativistic collisions is obtained by assuming that the
particles in the overlap zone of the two colliding nuclei are stopped,
while the spectators are propagating with their initial velocities. In the
original abrasion model \cite{kn:abr} the
excitation energy of the residues is calculated from the change of the
surface. For this geometrical abrasion process simple expressions for
the mass and energy of the participants and spectators have been given.

Gaimard and Schmidt \cite{kn:abr1} have improved this model by calculating the
excitation energy of the residue from the hole-state creation with
13.3 MeV excitation energy in average for each abrased particle:
\begin{eqnarray}
E^* = \alpha \cdot \Delta A, \quad  \alpha = 13.3 \mbox{MeV} .
\end{eqnarray}

\noindent
This prescription is in good agreement with the results of BUU
calculations \cite{kn:buu} , and yields the excitation energy of the
projectile independent of the target.  At small values of the mass loss
this picture underestimates the BUU prediction of the excitation energy,
and a higher value of $\alpha = 26.6$ MeV was suggested.
Also recent calculations of Campi et. al. \cite{kn:camp94} suggest somewhat
higher value for the excitation energy.
We have used both values in our calculations. The charge-to-mass ration
is assumed not to change by the abrasion process.

Furthermore we assume, that after the abrasion step the system equilibrates
fast. Estimating the equilibration times according to ref. \cite{kn:bert}, we
find indeed values which are small compared to the expansion times.

\subsection{Expansion}
\label{sec:eos}

Due to the supposed homogeneity in density, the radial flow is of
the form $\vec{v}(t,\vec{r}) = a(t) \vec{r}$ with $a=0$ initially.
 All the other variables -- the
density $\varrho(t)$, the temperature $T(t)$, the mass number $A(t)$ and the
charge $Z(t)$ depend only on time.

The evaporation
of nucleons yields the time evolution of mass and charge, and is
the only source for loss of energy and entropy.
The evaporation integral is calculated according to \cite{kn:he}, and
evaluated numerically. In this evaporation model
we need to know the mean free path not only at
nuclear matter density, but at all densities reached during the expansion.
We assume that the density dependence of the in-medium cross-section in
the range of $0.3 \varrho_0$ to $\varrho_0$ (normal nuclear density)
is essentially due to the
Pauli-blocking factor.
We found that the
contribution from the Pauli effect is inversely proportional to the density
for the temperatures, energies and densities in question.
Therefore, the mean free path is considered here as independ on the
density.

The equation of motion is determined from energy and entropy
conservations with loss terms which are due to the evaporation of nucleons.

The expansion is studied for a soft and a hard equation of state using,
respectively, the Skyrme forces SkM$^*$ and SIII, cf. \cite{kn:bra}.  We
follow the time evolution up to the turning point, where the density
reaches its minimum. The system spends most of its time around the
turning point, and hence the fragmentation process is expected to start
here. For too low excitations the system cannot enter the
$\varrho,T$ region of instability, and hence a heavy residue
will remain. For high excitations no turning point is encountered in the
expansion and, therefore, the system is expected to explode \cite{kn:rei}.

\begin{fig}{%
\iffigend
  {\epsfysize=17cm \epsfbox{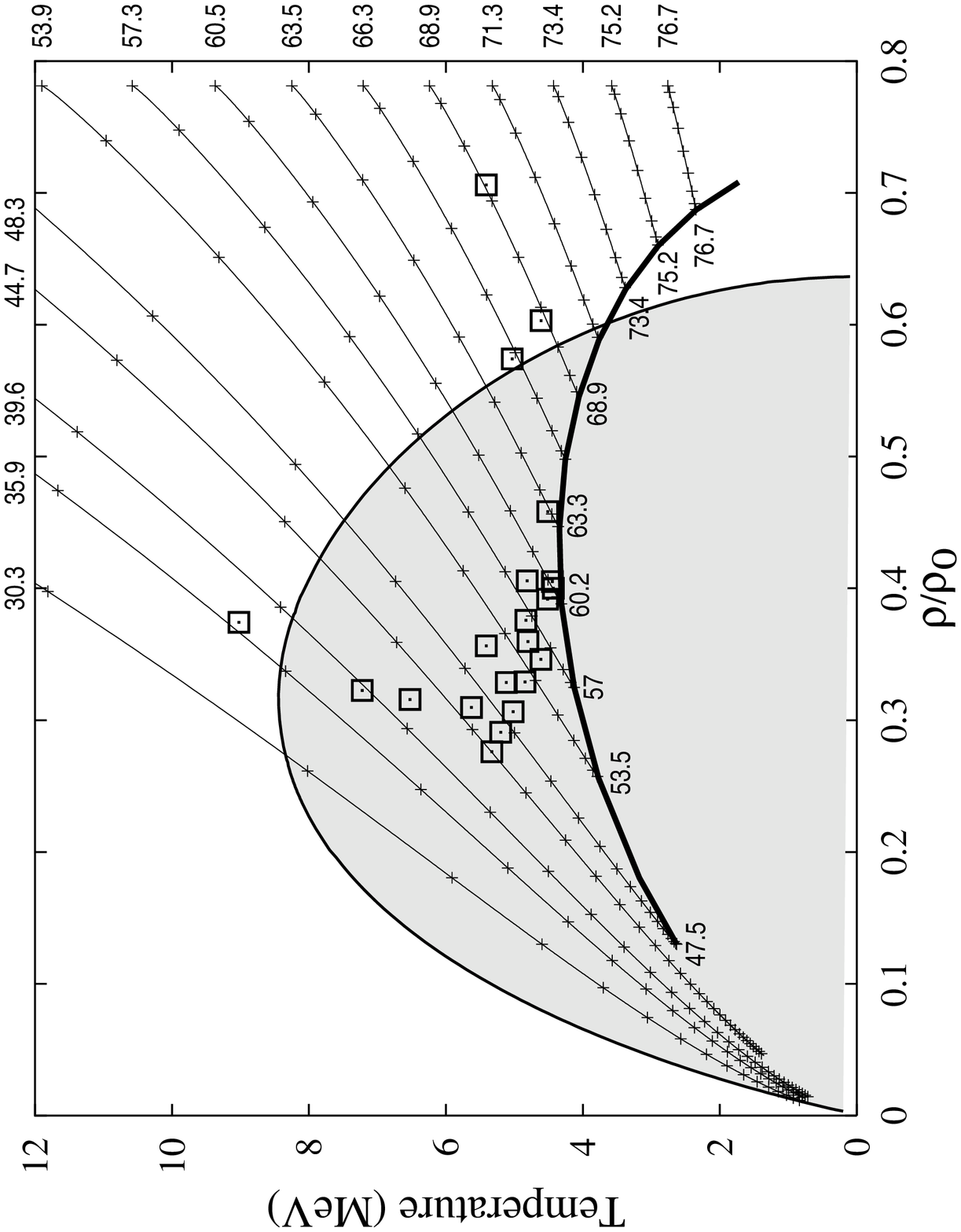}}
\else
  {\rotate[r]{\epsfxsize=9.5cm \epsfbox{soft.ps}}}
\fi
\caption{Time evolution in the ($T,\varrho$)-plane of the Au residue
for a soft equation of state (SkM$^*$). The
crosses on the trajectories indicate time steps of 5 fm/c. The numbers
give the charge numbers initially and  at the turning points which
are connected by the heavy solid line. The adiabatic spinodal region is
shadowed. The open boxes denote the experimental
data \protect\cite{kn:poch94} (without error bars).}
}
\label{fig-sof}
\end{fig}

\section{Results and Discussion}
\label{sec-res}

\begin{fig}{%
\iffigend
  {\epsfysize=17cm \epsfbox{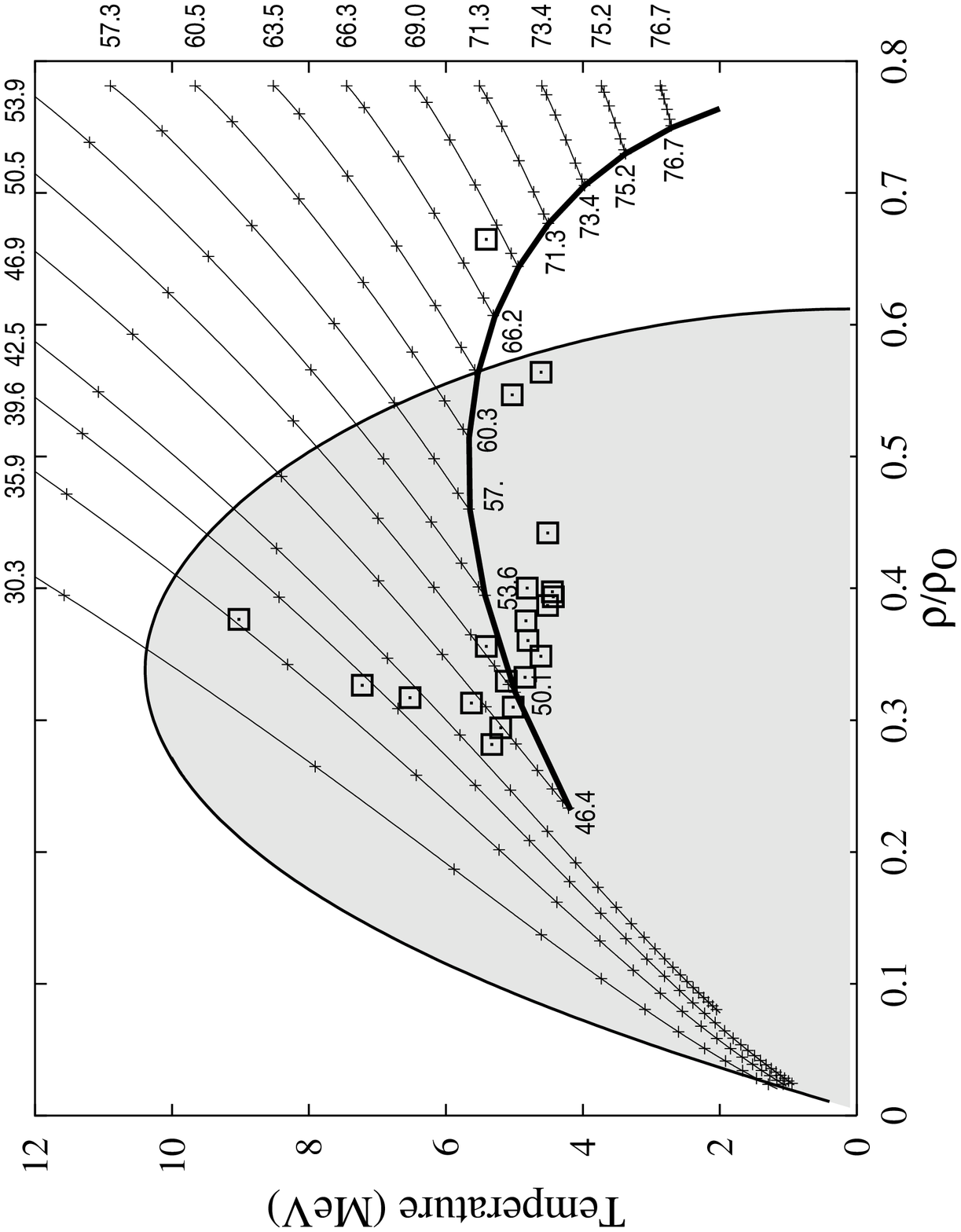}}
\else
  {\rotate[r]{\epsfxsize=9.5cm \epsfbox{hard.ps}}}
\fi
\caption{Same as Fig.~\protect\ref{fig-sof} for a stiff equation
of state (SIII)}
}
\label{fig-har}
\end{fig}

As an example of projectile multifragmentation we consider the reaction
Au (600 MeV/u -- 1 GeV/u) +X with X=C,Al,Cu which has been studied
experimentally by the ALADIN group \cite{kn:hub,kn:poch94}.

\subsection{Expansion trajectories}

According to the abrasion model (subsect.~2.1), the initial excitation energy
(temperature) is correlated with the charge number of the projectile residue.
These numbers are given in Figs.~\ref{fig-sof} and \ref{fig-har}.
The calculated expansion trajectories in the ($T,\varrho$)-plane are
significantly different for a soft (Fig.~\ref{fig-sof})  and a stiff
(Fig.~\ref{fig-har}) equation of state (EOS).
They deviate from adiabats due to the evaporation of one to four nucleons.
The number of evaporated nucleons is so small because of the fast cooling by
the expansion of the system.

We
follow the trajectories up to their turning points if reached within 200 fm/c.
The turning points are indicated by the heavy solid lines. Turning points exist
for initial temperatures up to about 14 MeV. We notice that the turning points
for the same initial excitation lie one to two MeV in temperature and
$0.1 \varrho_0$ to $0.2 \varrho_0$ in density higher for the stiff EOS as
compared to the soft one.
Moreover, we see that the turning points are located almost
at the
same temperatures independent of the initial excitation, i.e. around 4 MeV and
5 MeV for the soft and hard EOS, respectively. This plateau does not move
if one changes the initial excitation
energy (the $\alpha$ parameter in (1)), the only effect is, that the
charge numbers become different initially and at the turning points.

\subsection{Multifragmentation}

A natural criterion for multifragmentation to occur, is whether the system
reaches the region of volume instability (adiabatic spinodal region), where the
derivative of the pressure with respect to the density along an adiabat
becomes negative. Since the system is closed we consider the adiabatic process
to be the relevant one. In Figs.~\ref{fig-sof} to \ref{fig-sof15} the region
of instability
is shadowed and bounded by the spinodal. To enter the spinodal region and stay
there for more than 30 fm/c one
needs initial temperatures of about 8 MeV and 11 MeV for the soft and hard
EOS, respectively.

Because of the occurrence of turning points in the expansion,
the multifragmentation process of the projectile residue has a
unique feature:
the subsequent decay into fragments is expected to be rather free
from collective flow. This is different from the fragmentation of compressed
compound systems which are formed in central collisions.
As shown in Fig.~\ref{fig-sof15}, already for a small initial compression (1.5
$\varrho_0$) no turning points occur for realistic
initial excitation energies. Thus the fragmentation process becomes
qualitatively different and looks more like an explosion
\cite{kn:rei}.

\begin{fig}{%
\iffigend
  {\epsfysize=17cm \epsfbox{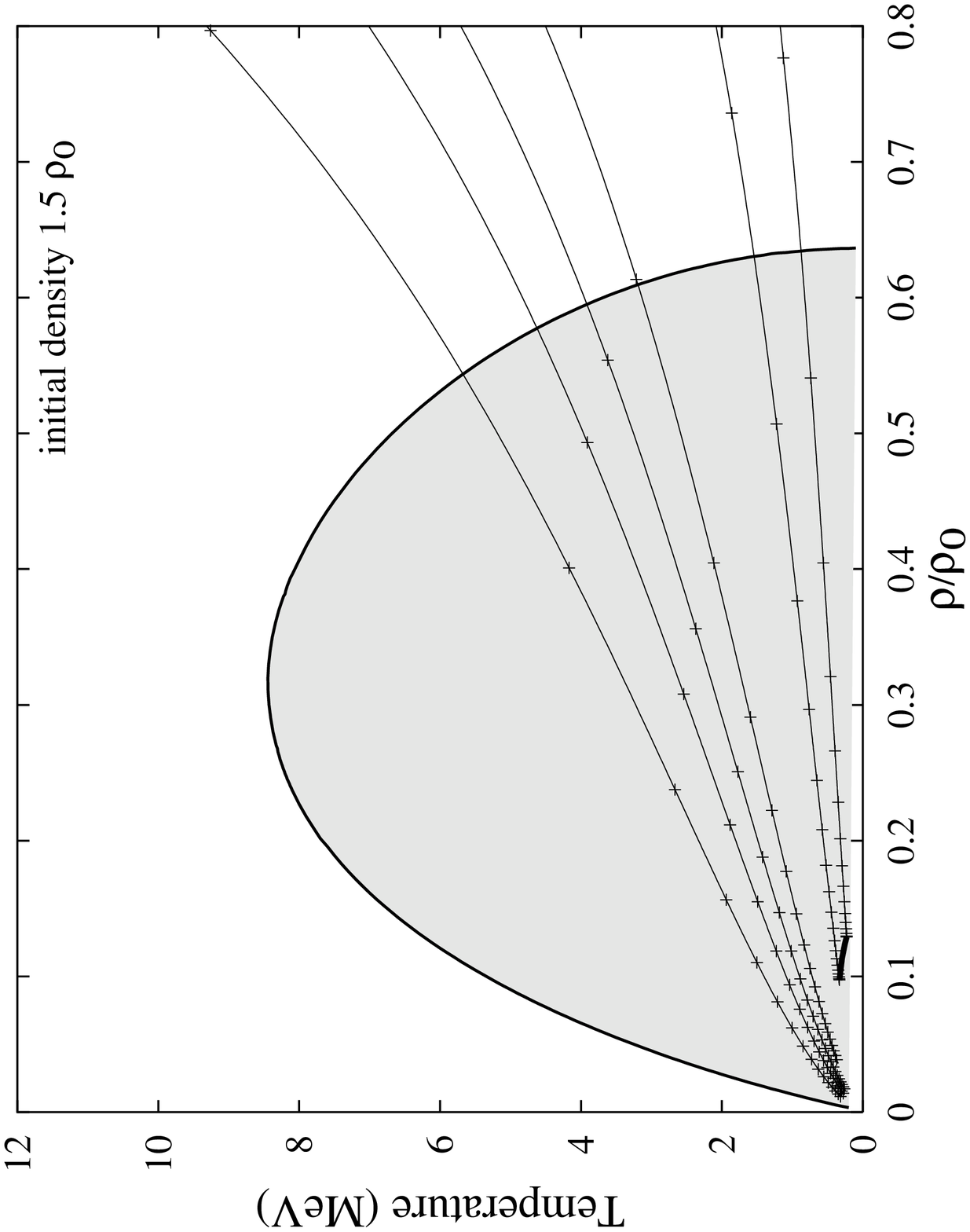}}
\else
  {\rotate[r]{\epsfxsize=9.5cm \epsfbox{soft15.ps}}}
\fi
\caption{Same as Fig.~\protect\ref{fig-sof} with additional
initial compression to 1.5 $\varrho_0$ for the soft EOS}
}
\label{fig-sof15}
\end{fig}

\subsection{Comparison with experimental results}

Pochodzalla et al.~\cite{kn:poch94} have determined the temperature of the
final break-up into fragments as function of the initial excitation energy. In
Figs.~\ref{fig-sof} and \ref{fig-har} we have plotted the experimental break-up
temperatures on the trajectories (or their extrapolations beyond the turning
points) for the corresponding initial temperature. The
experimental errors in the excitation energy transform into errors in the
density of about $\pm 0.06 \varrho_0$. We observe the following features.
\begin{itemize}
\item The onset of multifragmentation is around 8 MeV of initial temperature in
agreement with the soft EOS.
\item The experimental points, which are related to
initial temperatures between 6 MeV and 8 MeV, correspond to large values for
the maximum charge of the fragments, and hence suggest an evaporation-like
process.
\item For the stiff EOS no reasonable picture is obtained which would be
consistent with the experimental results.
\item There is some indication from molecular
dynamics~\cite{kn:pand85,kn:pand867} that
after a relatively fast expansion close to adiabats the system follows a path
close to $T=const$ with some tendency to increase the temperature.
Assuming that this is correct we would also conclude that the soft EOS
gives results which are consistent with experiment, whereas the stiff EOS does
not fit.
\item For initial temperatures larger than about 12 MeV to 14 MeV the system
reaches densities well below $0.3 \varrho_0$ before approaching the turning
point. At these densities
collisions between nucleons become rare, and hence this temperature is expected
to be observed in the final fragments.
\end{itemize}

\section{Summary and Conclusion}

We have studied the expansion of hot nuclei for a soft and a stiff equation of
state. Turning points are encountered for initial temperatures smaller than
about 14 MeV. Explosive events occurs for initial temperatures larger then 12
MeV to 14 MeV.

Projectile fragmentation appears to be the optimal process for the study of the
phase-transition region, because there is no compression involved in the
formation of initial state. Already small additional compression, as expected
in central collisions, lead to explosion.

The occurrence of turning points in
projectile fragmentation suggests to divide the multifragmentation process
into two steps. The first step is approximately described by the
expansion of a homogeneous nuclear drop, because the time necessary for the
development of instabilities is too large. The second step starts from the
turning point and leads to a relatively slow further expansion by developing
inhomogeneities. This slow evolution may be the reason why equilibrium
models~\cite{kn:gros,kn:bond,kn:camp94}  describe the fragment distribution
quite well.

Comparison with experimental results, obtained by the ALADIN
collaboration~\cite{kn:poch94}, gives evidence that the soft equation of state
is more realistic. In particular, the onset of multifragmentation at around 8
MeV initial temperatures is only consistent with the soft equation of state.
Furthermore, the break-up into fragments around a freeze-out
density of $\varrho/\varrho_0 = 0.3$ is indicated for initial temperatures
higher than about 12 MeV to 14 MeV.

We conclude that further dynamical studies of the fragmentation process and
comparison with experimental results may
yield more precise information about the equation of state in the low-density
region. In particular dynamical treatments of the fragmentation process
starting at the turning points are needed.

\ack

We would like to thank G. Bertsch, U. Lynen, W.F.J. M\"uller,
J. Pochodzalla and W. Trautmann for the fruitful discussions.

\end{document}